\pdfoutput=1

\documentclass[12pt]{article}
\usepackage{amsmath,epsf,amssymb,latexsym,cite,shadow,eucal,theorem,enumerate}
\usepackage[matrix,arrow,frame,import,curve,color]{xy}

\usepackage{graphicx}

\newtheorem{theorem}{Theorem}

{\theorembodyfont{\rmfamily}}

\newif\iffigs\figstrue

\DeclareFontFamily{U}{rsf}{}
\DeclareFontShape{U}{rsf}{m}{n}{
  <5> <6> rsfs5 <7> <8> <9> rsfs7 <10-> rsfs10}{}
\DeclareMathAlphabet\Scr{U}{rsf}{m}{n}

\def\O{\Scr{O}}
\def\C{{\mathbb C}}
\def\P{{\mathbb P}}

\def\Z{{\mathbb Z}}

\def\Ext{\operatorname{Ext}}
\def\Tor{\operatorname{Tor}}
\def\End{\operatorname{End}}

\def\Spec{\operatorname{Spec}}

\def\Gl{\operatorname{GL}}

\def\SU{\operatorname{SU}}
\def\GU{\operatorname{U{}}}

\def\Qb{\overline{Q}}
\def\nut{\widetilde{\nu}}
\def\qb{\overline{q}}
\def\ra{\rangle}
\def\gammab{\overline{\gamma}}
\def\phib{\overline{\phi}}
\def\p{\partial}

\def\CY{Calabi--Yau}
\def\LG{Landau--Ginzburg}

\def\cE{{\Scr E}}
\def\cF{{\Scr F}}

\def\ff#1#2{{\textstyle\frac{#1}{#2}}}

\def\poso#1{#1\save="x"!LD+<0pt,-0.5mm>;
  "x"!RD+<0pt,-0.5mm>**\dir{.}\restore}

\def\eqn#1#2{\begin{equation}#2
  \ifx{#1}{}\else\label{#1}\fi\end{equation}}

\begin{document}

\begin{titlepage}
\begin{flushright}
June 2011
\end{flushright}
\vspace{.5cm}
\begin{center}
\baselineskip=16pt
{\fontfamily{ptm}\selectfont\bfseries\huge
Elusive Worldsheet Instantons in\\
Heterotic String Compactifications\\[20mm]}
{\bf\large  Paul S.~Aspinwall and M.~Ronen Plesser
 } \\[7mm]

{\small

Center for Geometry and Theoretical Physics, 
  Box 90318 \\ Duke University, 
 Durham, NC 27708-0318 \\ \vspace{6pt}

 }

\end{center}

\begin{center}
{\bf Abstract}
\end{center}
We compute the spectrum of massless gauge singlets in some heterotic string
compactifications using Landau--Ginzburg, orbifold and non-linear
$\sigma$-model methods. This probes the worldsheet instanton
corrections to the quadratic terms in the spacetime superpotential. 
Previous results predict that some of these states remain massless when
instanton effects are included. We find vanishing masses in many 
cases not covered by these predictions.  However, we discover that in the case
of the $Z$-manifold the corrections do not vanish.  Despite this, in all the
examples studied, we find that the
massless spectrum in the  orbifold limit agrees with the nonlinear 
$\sigma$-model computation. 


\end{titlepage}

\vfil\break


\section{Introduction}    \label{s:intro}

The oldest approach to phenomenology in string theory is to compactify
the heterotic string on a \CY\ manifold $X$ together with a choice of
vector bundle $E\to X$.  Among compactifications with ${\cal N}=1$
spacetime supersymmetry, such models are unique in having a relatively
straightforward worldsheet formulation in terms of a $(0,2)$
superconformal field theory, and can be studied
beyond the supergravity approximation.  The effects of worldsheet
instantons in these models present some formidable technical
challenges and are far from completely understood.

In this paper we consider the moduli space of $(0,2)$-theories. Unlike
$(2,2)$-theories we may have obstructions to first order deformations
and, correspondingly, we have a notion of a superpotential in would-be moduli
fields. The superpotential is subject to instanton corrections at any
degree. A linear correction destabilizes the vacuum completely while a
quadratic correction removes a first-order deformation. Higher terms
in the superpotential affect the obstructions to the first-order deformations.

One may approach the computation of worldsheet instantons in two
ways. The direct way is to explicitly attack the geometry of the
rational curves in the target space. This has been pursued in works
such as \cite{DSWW:,Dist:res,Distler:1987ee,Berglund:1995yu,W:K3inst,
  Buchbinder:2002ic}. In this method one needs to compute Pfaffians
associated to each curve and then sum over all curves while correctly
taking into account multiple covers, etc.  In principle this allows a
complete computation of the full superpotential; in practice even the
quadratic terms present a daunting challenge.

The second approach, which avoids such formidable computations, is to
compare a perturbative non-linear $\sigma$-model count of massless
states with an exact conformal field theory method such as \LG\
theories. The difference will yield precisely the instanton
corrections to the quadratic terms in the superpotential.
Surprisingly, this difference is frequently zero.

The history of worldsheet instantons destabilizing the vacuum is
interesting. One knows that $(2,2)$ vacua are stable and so the
``standard embedding'' where $E$ is the tangent bundle $T(X)$ yields a
zero superpotential.
Beyond that it was expected that a generic CY 3-fold would have
rational curves giving generic instanton contributions leading to a
nonzero superpotential, indicating that the compactification specified
by $E$ is not in fact a supersymmetric vacuum
\cite{DSWW:}. Thus, anything other than a very specially chosen
vector bundle on a particular \CY\ manifold would fail to give a good
heterotic string compactification.

It was then realized \cite{Dist:res,Distler:1987ee} that the
contribution to worldsheet instanton corrections from a single
rational curve would be zero if the bundle $E$ split nontrivially over
the curve. While trivial splitting is generic, one could imagine
finding special examples where every rational curve had a nontrivial
splitting and thus the total instanton correction would be zero.

It was further realized that, in many cases which were understood as
an exact conformal field theory, even if single instantons contributed
nontrivially to the superpotential, there must a cancellation to
produce a zero net result \cite{Silverstein:1995re}. This was further
explored for the quintic threefold in \cite{Berglund:1995yu}.  This
apparently miraculous cancellation was explained in work by
Beasley and Witten \cite{Beasley:2003fx}. This latter paper also
indicated that such a cancellation must happen in general when $X$ is
a complete intersection in a toric variety $V$ and either
\begin{enumerate}
\item $E$ is a pullback of a vector bundle on $V$ or
\item $E$ is in the form of a monad naturally realized by the gauged
  linear $\sigma$-model.
\end{enumerate}
This accounts for a very wide range of possibilities and naturally one
might ask the question: {\em Can we ever measure nonzero instanton
  contributions to the superpotential from the exact conformal field theory?}

We will address this question in the easiest of settings, namely where
$E$ is the tangent bundle $T(X)$. The conformal field theory exhibits
$(2,2)$ worldsheet supersymmetry.  The model has an $E_6$ spacetime
gauge symmetry, and the spectrum includes massless $E_6$ singlet
chiral multiplets corresponding to first-order deformations of $T$.  
In the supergravity approximation, these are classically counted by
$h^1(\End(T))$, and perturbative corrections in $\alpha'$ to the
spacetime superpotential are prohibited by supersymmetry.  We further 
restrict our attention to terms in the superpotential quadratic in
these fields, which represent instanton-induced mass terms.

If worldsheet instantons contribute to these terms in the
superpotential then the number of these massless gauge singlet
chiral superfields, computed in 
the conformal field theory,  will be smaller than $h^1(\End(T))$. Thus we
may look for worldsheet instantons by comparing classical (i.e.,
perturbative nonlinear $\sigma$-model)
$h^1(\End(T))$ computations with exact results from the worldsheet.
In this paper we will use mirror symmetry to compute the exact
result. In particular, we will compute the spectrum from a
Landau--Ginzburg realization of the mirror.

Any deformation of $E$ that can be associated to a simple
deformation of the gauged linear $\sigma$-model Lagrangian
is a truly marginal deformation of the conformal
field theory, protected from worldsheet instanton-induced obstruction
at any degree by the
Beasley--Witten result. Thus we need to focus on more complicated
deformations.

We will see that the cancellation miracle persists in some cases
beyond Beasley-Witten (as had previously been observed in
\cite{meMP:singlets}). We will also find an example
where the number of massless singlets is drastically reduced from
$h^1(\End(T))$ and thus we do indeed have instanton corrections. This
example is given by the $Z$-manifold, i.e., a crepant resolution of
$T^6/\Z_3$.

Even though the $Z$-manifold suffers from instanton correction we will
see that it surprisingly obeys the ``heterotic McKay
correspondence''. That is, we may correctly compute $h^1(\End(T))$ by
counting untwisted singlets from a 6-torus and then adding in 27
copies of the twisted singlets from a $\Z_3$-quotient singularity as
explained in \cite{DHVW:}.  This is unexpected from the point of view
of string theory.  Unlike the McKay correspondence which, from the 
point of view of string theory, relates two computations of an
invariant quantity valid in different regions of the moduli space, the
quantity we are computing here -- the number of massless $E_6$ singlet
fields -- is not invariant and changes as we move about the moduli
space.  The calculation of \cite{DHVW:} is valid on the orbifold
locus, a nine-dimensional subspace describing the space with quotient
singularities (unresolved).  As we shall see, upon resolving the
singularities some of these modes acquire a mass.  $h^1(\End(T))$, on
the other hand, computes the number of fields whose masses vanish
exponentially as the volume of the space increases to infinity (in
such a way that the sizes of all holomorphic curves grow). It is
interesting that these numbers agree.


\section{Deformations of the Tangent Bundle} \label{s:tan}

Let $V$ be a toric variety and let $X\subset V$ be the desired \CY\
threefold given as a complete
intersection given by $s$ equations $f^a=0$, for $a=0,\ldots,s-1$.

We review the construction of $V$ to fix notation.  Let
$x_0,\ldots,x_{N-1}$ be the homogeneous coordinates on $V$. This is
the homogeneous coordinate ring in the sense of Cox \cite{Cox:}
\begin{equation}
  R = \C[x_0,\ldots,x_{N-1}].
\end{equation}

We have a short exact sequence
\begin{equation}
\xymatrix@1{
  0\ar[r]&M\ar[r]&\Z^{\oplus N}\ar[r]^\Phi& D\ar[r]&0,
} \label{eq:MZD}
\end{equation}
where $D$ is a lattice\footnote{Assumed to be torsion-free.} of rank
$r$. Each column of the matrix $\Phi$ can be thought of as a
$\GU(1)^r$ charge vector of the coordinates $x_i$. That is, $R$ has
the structure of an $r$-multigraded ring.

The toric variety is given as
\begin{equation}
  V = \frac{\Spec(R) - Z(B)}{(\C^*)^r}, \label{eq:V}
\end{equation}
where $B$ is the ``irrelevant ideal'' in $R$ and $Z(B)$ is the
associated subvariety of $\C^N$. $B$ is determined combinatorially
from the fan describing $V$.

Let $\mathbf{v}$ denote an element of the lattice $D$, i.e., an
$r$-vector. If $M$ is a multigraded $R$-module then we may shift
multi-gradings to form $M(\mathbf{v})$ in the usual
way. Correspondingly, if $\O_V$ is the structure sheaf of $V$, then we
may denote by $\O_V(\mathbf{v})$ the twisted sheaf associated to the
module $R(\mathbf{v})$. Line bundles on $V$ correspond to
$\O_V(\mathbf{v})$ for various $\mathbf{v}\in D$. If $V$ is smooth
then every element of $D$ defines a line bundle.

Let $\mathbf{q}_i$ denote the row vectors of the transpose of
$\Phi$. That is, $\mathbf{q}_i$ represents the multi-grading of the
homogeneous coordinate $x_i$. Let $T_V$ be the tangent sheaf of
$V$. Assuming $V$ is smooth, we have the generalization of the Euler
exact sequence for a toric variety \cite{BatCox:toric}
\begin{equation}
\xymatrix@1{
  0\ar[r]& \O_V^{\oplus r}\ar[r]^-{x_i\mathbf{q}_i}&
  \displaystyle{\bigoplus_{i=0}^{N-1}
  \O_V(\mathbf{q}_i)} \ar[r]& T_V\ar[r]&0.
} \label{eq:euler}
\end{equation}

Let
\begin{equation}
  \mathbf{Q} = \sum_{i=0}^{N-1} \mathbf{q}_i.
\end{equation}
For the complete intersection $X$ we have the adjunction exact
sequence:
\begin{equation}
\xymatrix@1{
  0\ar[r]&T_X\ar[r]&T_{V|X}\ar[r]&\bigoplus_a\O_X(\mathbf{Q}_a)\ar[r]&0,
} \label{eq:adj}
\end{equation}
where $\mathbf{Q}_a$ is the multi-degree of the equation $f^a$. The
\CY\ condition is $\sum_a\mathbf{Q}_a=\mathbf{Q}$.

Since all the sheaves in (\ref{eq:euler}) are locally-free, we may
restrict to $X$ and the sequence will remain exact. Combining this
with the sequence (\ref{eq:adj}) yields the following fact. The
tangent sheaf is given by the cohomology at the middle term of
\begin{equation}
\xymatrix@1{
  0\ar[r]& \O_X^{\oplus r}\ar[r]^-{x_i\mathbf{q}_i}&\displaystyle{\bigoplus_i^{\phantom{N}}
  \O_X(\mathbf{q}_i)} \ar[r]^-{\partial_iW_a}&\displaystyle{\bigoplus_a^{\phantom{N}} 
  \O_X(\mathbf{Q}_a)}\ar[r]&0.
} \label{eq:TXci}
\end{equation}

Obviously we may deform this complex to
\begin{equation}
\xymatrix@1{
  0\ar[r]& \O_X^{\oplus r}\ar[r]^-{E}&\displaystyle{\bigoplus_i^{\phantom{N}}
  \O_X(\mathbf{q}_i)} \ar[r]^-{J}&\displaystyle{\bigoplus_a^{\phantom{N}} 
  \O_X(\mathbf{Q}_a)}\ar[r]&0,
} \label{eq:TXdef}
\end{equation}
for generic matrices $E$ and $J$ of the correct multi-degree. By
varying $E$ and $J$ we produce a family of sheaves containing the
tangent sheaf. This family of sheaves can be understood in terms of
the gauged linear $\sigma$-model \cite{W:phase,Kreuzer:2010ph}. As such these
are deformations of the $(0,2)$-model protected by Beasley--Witten.
In general this is a subspace of the moduli space of $(0,2)$
deformations.  Correspondingly, there are deformations of $T_X$ that
this description does not capture.  These deformations are not
protected and worldsheet instanton effects can be nontrivial.

We thus need a more complete description of the space of deformations
of $T_X$.  
The first order deformations of $T_X$ are given by
the vector space
\begin{equation}
   \Ext^1(T_X,T_X) = H^1(X,\End(T)).
\end{equation}
We may follow \cite{meMP:singlets} and compute this as follows. The
sheaf $\End(T)$ can be written as the cohomology of (\ref{eq:TXci})
tensored with its dual. That is,

\begin{multline}
\xymatrix@1{
\bigoplus_a\O_X(-\mathbf{Q}_a)^{\oplus r}\ar[r]&
{\begin{matrix}\bigoplus_{i,a}\O_X(\mathbf{q}_i-\mathbf{Q}_a)\\
  \oplus\\
  \bigoplus_i\O_X(-\mathbf{q}_i)^{\oplus r}\end{matrix}}\ar[r]&
\poso{{\begin{matrix} \O_X^{\oplus r^2}\\
\oplus\\
\bigoplus_{i,j}\O_X(\mathbf{q}_i-\mathbf{q}_j)\\
\oplus\\
\O_X\end{matrix}}}\ar[r]&}\\[10mm]
\xymatrix@1{
{\begin{matrix}\bigoplus_i\O_X(\mathbf{q}_i)^{\oplus r}\\
  \oplus\\
  \bigoplus_{i,a}\O_X(\mathbf{Q}_a-\mathbf{q}_i)\end{matrix}}\ar[r]&
  \bigoplus_a\O_X(\mathbf{Q}_a)^{\oplus r}&\hbox{\phantom{ww}}
} \label{eq:HomTT}
\end{multline}
The cohomology is at the term we have underlined. We will consider the
underlined term position ``zero'' in the complex.
If a sheaf is
presented as the cohomology (at position zero) of a complex 
\begin{equation}
\xymatrix@1{
  \ldots\ar[r]&\cE^{-1}\ar[r]&\cE^{0}\ar[r]&\cE^{1}\ar[r]&\ldots,
}
\end{equation}
then there is a spectral sequence \cite{meMP:singlets} converging to
the cohomology of the sheaf whose $E_1$ term is given by
\begin{equation}
  E_1^{p,q} = H^q(X,\cE^p).
\end{equation}

As explained in \cite{meMP:singlets}, there is a strong resemblance
between row 0 (i.e., $q=0$) and the first order deformations of
the linear $\sigma$-model (\ref{eq:TXdef}). The space
$H^0(V,\O_V(\mathbf{q}))$ is the space of global sections of 
$\O_V(\mathbf{q})$ and its dimension is counted by the number of
monomials in $R$ with multi-degree $\mathbf{q}$. We can count the
deformations of (\ref{eq:TXdef}) by considering all possible matrices
of polynomials $E$ and $J$ such that $J.E=0$ and then subtracting the
number of reparametrizations induced by changes of homogeneous
coordinates. This amounts to computing the cohomology in the middle
term of the sequence:
\begin{equation}
\xymatrix@1{
{\begin{matrix} H^0(\O_V^{\oplus r^2})\\
\oplus\\
\bigoplus_{i,j}H^0(\O_V(\mathbf{q}_i-\mathbf{q}_j))\\
\oplus\\
H^0(\O_V)\end{matrix}}\ar[r]&
{\begin{matrix}\bigoplus_iH^0(\O_V(\mathbf{q}_i)^{\oplus r})\\
  \oplus\\
  \bigoplus_iH^0(\O_V(\mathbf{Q}-\mathbf{q}_i))\end{matrix}}\ar[r]&
H^0(\O_V(\mathbf{Q})^{\oplus r}) .
} \label{eq:GLSMEJ}
\end{equation}
Comparing this to the zeroth cohomology of (\ref{eq:HomTT}) we see two
obvious differences:
\begin{enumerate}
\item The computation is on $V$ rather than $X$ and
\item The left two terms of (\ref{eq:HomTT}) are missing.
\end{enumerate}
In simple cases these differences have no effect and the zeroth row
of the spectral sequence accurately represents the deformations as seen
by the linear $\sigma$-model.

For the quintic threefold in $\P^4$ this is the complete story. There
is no contribution to $h^1(\End(T))$ from any rows other than the
zeroth row. Thus all deformations of $T$ are understood in terms of
the linear $\sigma$-model. They are unobstructed and cannot be spoiled by
worldsheet instantons.

\section{The Octic} \label{s:8ic}

Let us summarize the results of \cite{meMP:singlets}
where $X$ is the resolved octic hypersurface in
$\P^4_{\{2,2,2,1,1\}}$.
\begin{itemize}
\item There are 179 deformations of $T$ that are seen by the gauged
  linear $\sigma$-model and are given by the bottom row of the
  spectral sequence.
\item The total count of $h^1(\End(T))$ depends on the complex
  structure. Generically it is 188 while for the Fermat hypersurface it is
  200. All even values between these extremes can be obtained by choosing
  suitable octic defining equations.
\item For generic values of the map $E$ the number of deformations is
  188. Thus some, if not all, of the extra 12 deformations associated
  to the Fermat complex structure are obstructed.
\end{itemize}

This jumping of the value of $h^1(\End(T))$ shows that we have a
nontrivial superpotential for the singlets.

At the Gepner point we have 206 singlets associated with the bundle
data. (That is 206 singlets aside from deformations of complex
structure, deformations of K\"ahler form and partners of the
$\GU(1)^4$ gauge symmetry.) By deforming the superpotential of the
{\em mirror\/} Landau--Ginzburg theory we lose 6 of these 206. Thus 6
singlets are an artifact of being at the Gepner radius.

On deforming the superpotential of the \LG\ theory we can lose up to
12 more singlets. In fact we get {\em perfect\/} agreement between the \LG\
theory and the geometrical result. For any defining equation, i.e.,
superpotential, the number of computed singlets associated to
$h^1(\End(T))$ is between 188 and 200 and the \LG\ matches the
geometry.\footnote{Note that this \LG\ computation is stuck at small radius while the
classical computation is stuck at large radius. It is just conceivable that
this perfect agreement at large and small radius is spoiled at intermediate
radii. We will assume this is not the case.}

One should be able to, in principle, compute the precise obstruction
theory for the bundle and compute all correlation functions between
the singlets in the \LG\ model. This would allow a precise comparison
of the superpotential between geometry and the exact result. This, in
turn, would show if there were any instanton corrections to the
superpotential. We have not done this but the agreement above does show
that there are no corrections that would affect the {\em masses\/} of
the singlets.

This shows that there is some ``miracle'' that kills instanton
corrections to these $(0,2)$-models that goes beyond
\cite{Silverstein:1995re,Beasley:2003fx}. Indeed we have checked many
examples of \CY\ {\em hypersurfaces\/} in toric varieties and we always find
agreement of singlet counting between classical and exact methods.

It is natural to ask, therefore, whether this unnatural agreement
persists for all $(0,2)$-models with the standard embedding. We will
see that it is not the case.

\section{The Z-Manifold}

\subsection{Geometry}
Let $T$ be the 2-torus given by $z\in\C$ under the identification
$z\mapsto z+1$ and $z\mapsto z+\omega$, where $\omega=\exp(2\pi
i/3)$. Let $Z_0$ be the orbifold $(T\times T\times T)/\Z_3$ where the
$\Z_3$ is generated by the action
\begin{equation}
   g:(z_0,z_1,z_2)\mapsto (\omega z_0,\omega z_1,\omega z_2).
\end{equation}
This is the well-known $Z$-orbifold introduced in \cite{DHVW:} in
which the associated conformal field theory was written in terms of
free fields and the spectrum computed exactly.

The $\Z_3$ action has 27 fixed points yielding 27 singularities in
$Z_0$. These can be resolved by blowing up each point with a $\P^2$
exceptional set to yield the $Z$-manifold. 

It is very easy to compute the Hodge numbers of $Z$ by using
homology. Each blow-up introduces a 4-cycle in terms of the
exceptional $\P^2$. Adding this 27 to the 9 invariant cycles from the
covering 6-torus yields $b_2=h^{1,1}=36$. Similarly one can argue that
$h^{2,1}=0$ and the $Z$-manifold is rigid.

It is considerably harder to compute $h^1(\End(T))$. String theory
implies there is a way to add local contributions of the blow-ups to
some global contribution of the 6-torus but we do not know how to {\em
  rigorously\/} formulate this geometrically.\footnote{An ALE space
  must be deformed before it is glued in and the torus metric must be
  deformed away from being flat. Such deformations may {\em a priori\/}
  affect $h^1(\End(T))$.}  Instead we use another construction of
the $Z$-manifold from which we do know how to extract $h^1(\End(T))$.

Since toric varieties offer a tractable path, we embed $Z$ into a
toric variety. Unfortunately, it cannot be embedded as a hypersurface
since all such hypersurfaces have a mirror \cite{Bat:m}. We can,
however write it as the complete intersection of 3 equations.

The computation of $h^1(\End(T))$ for the $Z$-manifold is very lengthy
and we confine the details to an appendix. The result is that
\begin{equation}
  h^1(\End(T)) = 208.
\end{equation}

It turns out that, of these 208, only 6 come from the bottom row of the
spectral sequence and are thus protected by Beasley--Witten from
instanton corrections.

\subsection{The Landau--Ginzburg Picture}

At a point in the $(2,2)$ moduli space, the superconformal nonlinear
sigma model on the $Z$-manifold is equivalent to a
Gepner model \cite{Gepner:1987qi} .  This is most directly seen by recalling that the
sigma model on an elliptic curve, at precisely the complex structure
exhibiting a $\Z_3$ symmetry mentioned in the previous
section, is equivalent at one point in its K\"ahler moduli space to 
a Gepner model, a $\Z_3$ quotient of the product $A_1^{\otimes 3}$.
The construction of the $Z$ as an orbifold then shows that at a point
in the nine-dimensional moduli space of the orbifold theory (before
blowing up) the model is equivalent to a $\left(\Z_3\right)^4$
quotient of the product of superconformal minimal models 
$A_1^{\otimes 9}$. 

In this form, it is straightforward to count the 270 massless $E_6$
singlet fields at this point in the moduli space \cite{Lutken:1988hc}.
Of these, of course, 36 are the $(2,2)$ moduli, but this still leaves
234 singlets, more than the 208 found at large radius above.  One part
of the discrepancy is clear.  At the Gepner point the theory exhibits
a $\GU(1)^6\times\SU(3)$ gauge symmetry, which is broken by generic
(K\"ahler) deformations.  This leads to $D$-term masses (via the Higgs
mechanism) for 14 of the singlet fields.  Masses for any of the
remaining 220 fields are generated by the spacetime superpotential.

In the Gepner model mentioned (and the associated \LG\ model) the
$(2,2)$ moduli are all twisted fields under the orbifold projection,
making it difficult to study the model away from this one point.  To
get around this we study instead the mirror of the $Z$-manifold.  At a
point in its moduli space, the superconformal theory on this is
equivalent to a quotient of the same Gepner model \cite{GP:orb}.  The
construction leads to a $\Z_3^5$ quotient of the product of minimal
models.  It is simpler, in this case, to note that since an elliptic
curve is its own mirror, up to a relabeling of the fields this is
equivalent to a $\Z_3^2$ quotient.  As expected, the mirror model has
a 36-dimensional space of $(2,2)$ deformations.  Of these, 30 are untwisted
under the quotient and can be represented in the associated \LG\ model
as deformations of the worldsheet superpotential.  The methods of
\cite{Kachru:1993pg}\ then enable a computation of the number of
 massless singlet fields at any point in this 30-dimensional subspace
 of the full moduli space.  The orbifold locus intersects this
 subspace along a three-dimensional subspace.

The \LG\ model contains nine $(2,2)$ chiral superfields which we 
denote $X^i, Y^i, Z^i,\ i = 1\ldots 3$ interacting via a cubic
superpotential.  The model has a $\GU(1)$ R-symmetry under which 
the fields all have charge $1/3$.  In the infrared this model flows to 
a $(2,2)$ superconformal field theory with a $\GU(1)\times\GU(1)$ R-symmetry.
The Gepner model (after GSO projection) is a
$\Z_6$ orbifold of this, and our mirror model is a further quotient by
a $\Z_3$ generated by $(X^i,Y^j,Z^k)\mapsto(\omega X^i, \omega^2 Y^j, Z^k)$.  The
most general invariant superpotential can be written as
\begin{equation}
W = A_{ijk} X^i X^j X^k + B_{ijk} Y^i Y^j Y^k + C_{ijk} Z^i Z^j Z^k +
D_{ijk} X^i Y^j Z^k\ . 
\end{equation}
The 57 parameters in this superpotential are subject to an action of 
$\Gl(3)^3$, which allows us to bring it to the form 
\begin{multline}
W = \sum (X^i)^3 - 3 A_x X^1 X^2 X^3 + \sum (Y^i)^3 - 3 A_y Y^1 Y^2 Y^3\\ + \sum
(Z^i)^3 - 3 A_z Z^1 Z^2 Z^3  + D_{ijk} X^i Y^j Z^k\ .
\end{multline}
This is  our explicit representation of the 30-dimensional family.  
The orbifold locus corresponds to $D_{ijk}=0$.

We can consider $(0,2)$ deformations of this as a \LG\ model; this is
a degenerate version ($E=0$) of the gauged linear sigma model counting
(\ref{eq:TXdef}) and we find 82 such deformations, providing a lower
bound on the number of  massless singlets at any point in the moduli space.

Supersymmetric ground states are found using the 
left-moving $N=2$ superconformal algebra in the cohomology of the
right-moving supercharge $\Qb$ and classified by their charges 
$(q,\bar q)$ under the $\GU(1)$ symmetry contained in this algebra and 
under the right-moving $\GU(1)$ symmetry inherited from the $(2,2)$ 
structure.  States are described as excitations by the lowest
oscillator modes of free
bosonic fields $\phi_a^i$ (labeled by $a=x,y,z$ and $i=1,2,3$) with 
charges $(1/3,1/3)$ and left-moving fermionic fields $\gamma_a^i$ with 
charges $(-2/3,1/3)$.

The generator of the Gepner
quotient acts as $e^{\pi i q}$ while the additional $\Z_3$ quotient
acts as $\phi_a^i\to\omega^{w_a}\phi_a^i;\
\gamma_a^i\to\omega^{w_a}\gamma_a^i$ with $w = (1,-1,0)$.  Twisted
vacua are labeled by $(k;l)$ with $k=0,\ldots, 5$ and $l=0,1,2$.  
It should be emphasized that this orbifold construction of the mirror
of the $Z$-manifold is distinct from the orbifold construction of the
$Z$-orbifold itself as in \cite{DHVW:}.  In particular twisted states
in one orbifold need not correspond to twisted states in the other.
In the $(k;l)$ sector the fields have a twisted moding
\begin{align}
\phi_a^i(z) &= \sum_{s\in \Z-\nu_a} x_a^i{}_{(s)} z^{-s-1/6},      
&\qquad
& \gamma_a^i(z)  = \sum_{s\in \Z-\nut_a} \gamma_a^i{}_{(s)} z^{-s-2/3} \nonumber\\
2\p\phib_{ai}(z) &= \sum_{s\in\Z+\nu_a} \rho^{}_{ai}{}_{(s)} z^{-s-5/6}, 
&\qquad 
&\gammab_{ai}(z) = \sum_{s\in\Z+\nut_a} \gammab_{ai}{}_{(s)} z^{-s-5/6},
\end{align}
where 
\begin{align}
\nu_a(k;l) &= {k\over 6} + {l w_a\over 3} \pmod 1 &\qquad
0&\le\nu_a<1\nonumber\\
\nut_a(k;l) &= {-k\over 3} + {l w_a\over 3} \pmod 1 &\qquad
-1&<\tilde\nu_a\le 0
\end{align}
The ground state energy and charges of the twisted vacua are 
\begin{eqnarray}
E(k;l) &=& -{5\over 8} + {3\over 2}\sum_a \Bigl( \nu_a(1-\nu_a)
  +\nut_a(1+\nut_a) \Bigr)\quad k\ {\rm odd}\nonumber\\
q(k;l) &=& -\sum_a\left( 2 \nut_a + \nu_a +\ff{1}{2}   \right)\nonumber\\
\qb(k;l) &=& \sum_a \left(  \nut_a  +2\nu_a - \ff{1}{2}\right)\ .\\
\end{eqnarray}
For $k$ even $E(k;l)=0$.

Massless fermions arise in R (odd $k$) sectors and correspond to
excitations with $E=0$.  $E_6$  singlets are characterized by $q=0$.
The right-moving charge $\qb$ of a state determines the
spacetime multiplet to which the fermion belongs: $\qb = \pm\ff32$ are
fermions in vector multiplets while $\qb=\pm\ff12$ are fermions in 
chiral multiplets.  We construct states with $E=q=0$ by acting on
$|k;l\ra$ with the lowest excited modes, which we denote 
\begin{equation}
x_a^i \equiv x_a^i{}_{(-\nu_a)}, \quad \rho_{ai} \equiv \rho^{}_{ai}{}_{(\nu_a-1)}, 
\quad \gamma_a^i \equiv \gamma_a^i{}_{(-1-\nut_a)}, \quad
\gammab_{ai} \equiv \gammab_{ai}{}_{(\nut_a)}.
\end{equation}
In describing the $\Qb$ cohomology we will also need the conjugate modes
\begin{equation}
x^\dag{}_{ai} \equiv \rho^{}_{ai}{}_{(\nu_a)}, \quad \rho^\dag{}_a^i \equiv x_a^i{}_{(1-\nu_a)}, \quad
\gamma^{\dag}{}_{ai} \equiv \gammab_{ai}{}_{(1+\nut_a)}, \quad
\gammab^{\dag}{}_a^i \equiv \gamma_a^i{}_{(-\nut_a)}.
\end{equation}
Acting on these states, the $\Qb$ operator takes the general form
\begin{equation}\label{eq:LGqbar}
\Qb = \sum_{a,i} \left\{ \gamma_a^i \p_{ai}W_{1+\nut_i} + \gammab^\dag{}_{ai}\p_{ai}W_{\nut_i}\right\}.
\end{equation}

The quotient breaks the $S_9$ permutation symmetry of the product to a
subgroup.  Of interest to us is an unbroken $S_3$ subgroup permuting
the $a$ indices of the fields.  This relabels the generators of the
quotient group and correspondingly permutes the twisted sectors.

\subsubsection{Untwisted States}

In the untwisted R sector $(k;l) = (1;0)$ we have $\nu_a=\ff16,\
\nut_a=-\ff13$ so that the ground state has $E(1;0) = -1,\ 
q(1;0) = 0,\ \qb(1;0) = -\ff32$.  The complex of $E=q=0$
states upon which $\Qb$ acts is 
\begin{equation}
\xymatrix@C=20mm@R=5mm{
  \bar q=-\ff32&\bar q=-\ff12&\\
{\begin{matrix}
\phi_a^i\rho_{aj}|1;0\ra_{27}\\
\oplus\\
\gammab_{ai}\gamma_a^j|1;0\ra_{27}
\end{matrix}}
\ar[r]^-\Qb&
{\begin{matrix}
\phi_a^i \phi_a^j\gamma_a^k|1;0\ra_{54}\\
\oplus\\
\phi_a^i \phi_b^j\gamma_c^k|1;0\ra_{81}\\
\end{matrix}}
&{\begin{matrix}\\\\(a\ne b\ne c)\end{matrix}}
}
\end{equation}
where subscripts on kets indicate the dimension of the space of states
of a given form; repeated indices on fields are not summed.
We write $\Qb$ as a sum of three terms 
\begin{equation}
\Qb = \Qb_G + \Qb_O + \Qb_D\ .
\end{equation}
$\Qb_G$, the supercharge at the Gepner point, is in this sector given
by
\begin{equation}
\Qb_G = 6\sum_{a,i}\gamma_a^i x_a^i \rho^\dag{}_a^i +
3\gammab^\dag{}_a^i \left(x_a^i\right)^2\ .
\end{equation}
This has a nine-dimensional kernel spanned by
$J_a^i = \ff13\left(x_a^i\rho_{ai} - 2\gammab_{ai}\gamma_a^i\right)|1;0\ra$,
indicating that the enhanced gauge group at the Gepner point has rank
eight (one generator is the $\GU(1)\subset E_6$), 
and that the untwisted sector gives rise to 90 chiral singlets.
$\Qb_O$, the additional charge at a generic point on the orbifold
locus $D=0$ is in this sector given by 
\begin{equation}
\Qb_O = -3\sum_aA_a\sum_{i\ne j\ne k}
\left[\gamma_a^i(x_a^j\rho^\dag{}_a^k + x_a^k\rho^\dag{}_a^j ) +
  \gammab^\dag{}_a^ix_a^j x_a^k\right]\ .
\end{equation}
Adding this to $\Qb_G$ with generic $A_a$ reduces the dimension of the kernel to three,
spanned by $J_a=\sum_i  J_a^i$, indicating an
enhanced gauge group of rank two, and a total of 84 chiral singlets.
Adding 
\begin{align}
\Qb_D =& \sum_{ijk} D_{ijk} \left[\gamma_x^i(y^j\rho^\dag{}_z^k + z^k\rho^\dag{}_y^j ) +
  \gammab^\dag{}_x^iy^j z^k\right.\nonumber\\
&+ 
\gamma_y^j(x^i\rho^\dag{}_z^k + z^k\rho^\dag{}_x^i ) +
  \gammab^\dag{}_y^jx^i z^k + 
\left.\gamma_z^k(y^j\rho^\dag{}_x^i + x^i\rho^\dag{}_y^j ) +
  \gammab^\dag{}_z^kx^iy^j\right]
\end{align}
leaves a one-dimensional kernel (guaranteed by the quasihomogeneity of
$W$) generated by $J = \sum_a J_a$, 
indicating the gauge symmetry is reduced to $E_6$ and leaving 82
neutral chiral multiplets.

\subsubsection{Twisted States}

Since the $S_3$ symmetry permutes the twisted sectors, a calculation
in any one is sufficient to produce the entire spectrum.  In the
$(k;l) = (1;1)$ sector we have $\nu(1;1) = (\ff12,\ff56,\ff16),\ \nut(1;1) = 
(0,-\ff23,-\ff13)$ and hence $E(1;1) = -\ff12,\ q(1;1) = -1,\ \qb(1;1) =
\ff12$.  The complex of $E=q=0$ states is 
 \begin{equation}
\xymatrix@C=10mm@R=5mm{
  \bar q=-\ff32&\bar q=-\ff12&\bar q=\ff12&\qb=\ff32\\
\rho_y^3\gammab_x^3|1;1\ra_{10}
\ar[r]^-\Qb&
{\begin{matrix}
\rho_x\gammab_x^2|1;1\ra_{9}\\\oplus\\
z\rho_y^2\gammab_x^2|1;1\ra_{54}\\\oplus\\
\rho_y\gammab_x\gammab_z|1;1\ra_{27}\\\oplus\\
\rho_y\gammab_x^3\gamma_y|1;1\ra_{9}
\end{matrix}}
\ar[r]^-\Qb&
{\begin{matrix}
x\gammab_x|1;1\ra_{9}\\\oplus\\
z^2\rho_y\gammab_x|1;1\ra_{54}\\\oplus\\
z\gammab_z|1;1\ra_{9}\\\oplus\\
z\gammab_x^2\gamma_y|1;1\ra_{27}
\end{matrix}}
\ar[r]^-\Qb&
z^3|1;1\ra_{10}
}
\end{equation}

In this sector we have 
\begin{equation}
\Qb_G = 3\sum_i\left[\gamma_x^i \left(\rho^\dag{}_{xi}\right)^2 +
  2\gammab^\dag{}_x^ix^i\rho^\dag{}_x^i
+ \gamma_y^i \left(\rho^\dag{}_{yi}\right)^2 +
2\gammab^\dag{}_y^iy^i\rho^\dag{}_y^i
+ 2\gamma_z^i z^i\rho^\dag{}_{zi} +
\gammab^\dag{}_y^i\left(z^i\right)^2\right]\ .
\end{equation}
Acting on the $\qb=\ff32$ space this has a one-dimensional kernel
generated by $\rho_{y1}\rho_{y2}\rho_{y3}\gammab_x^3|1;1\ra$; the
six vector multiplets arising from twisted sectors fill out the
enhanced gauge group $\GU(1)^6\times \SU(3)$ with the Cartan
subgroup generated by the charges associated to $J_a-\ff13 J$. 

Acting on the $\qb=\ff12$ space we find a 39-dimensional kernel, so that
each twisted sector contributes 30 massless $E_6$ singlet chiral
multiplets; when added to the 90 we found in the untwisted sector this
reproduces the expected 270 singlets at the Gepner point.  

The analysis is made easier by organizing the zero-energy states into
$\SU(3)$ multiplets and recalling that $\Qb_G$ as well as $\Qb_O$ are 
$\SU(3)$ invariant and that $\Qb$ acts within a given twist sector.  We find 
\begin{equation}
\xymatrix@C=10mm@R=5mm{
  \bar q=-\ff32&\bar q=-\ff12&\bar q=\ff12&\qb=\ff32\\
{\begin{matrix}
{\bf 1}^{34}\\
\oplus\\
{\bf 8}^{10}\\
\\
\\
\end{matrix}}
\ar[r]^-\Qb&
{\begin{matrix}
{\bf 1}^{36}\oplus{\bf 3}^{27}\\
\oplus\\
{\bf 8}^9\\
\oplus\\
{\bf 1}^{54}\oplus{\bf 3}^{108}\oplus{\bf{\overline 3}}^{54}
\end{matrix}}
\ar[r]^-\Qb&
{\begin{matrix}
{\bf 1}^{36}\oplus{\bf{\overline 3}}^{27}\\
\oplus\\
{\bf 8}^9\\
\oplus\\
{\bf 1}^{54}\oplus{\bf 3}^{54}\oplus{\bf{\overline 3}}^{108}
\end{matrix}}
\ar[r]^-\Qb&
{\begin{matrix}
{\bf 1}^{34}\\
\oplus\\
{\bf 8}^{10}\\
\\
\\
\end{matrix}}
}
\end{equation}
where the first row represents the contribution of the untwisted sectors
and the last the contributions of the twisted sectors.  Adjoints, in
the second row, are
counted separately as they are assembled from twisted and untwisted
states.  In the untwisted sector we computed above that $\Qb_G$ has
rank 27 when acting on the space of $\SU(3)$ singlets.  Acting on the
adjoints it has the maximum rank possible.  
Since untwisted states with $\qb>0$ arise in the
conjugate $(5;0)$ sector the map $\qb=-\ff12\to\qb=\ff12$ vanishes for
untwisted states.  In the twisted sector,
acting on the $\SU(3)$ singlet states $\Qb_G$ has rank 36.  In the
$(1;1)$ sector the three-dimensional kernel at $\qb=-\ff12$ is spanned
by $\rho_{xi}\gammab_{xj}\gammab_{xk}|1;1\ra$ with $i\ne j\ne k$.  The rank
on the $\SU(3)$ charged fields is maximal and we find that the
physical states are vector multiplets in the representation
 \begin{equation}
{\bf 1}^7\oplus{\bf 8},
\end{equation}
of which one of the singlets is a Cartan generator of $E_6$, 
and chiral multiplets in the representation
\begin{equation}
\underbrace{{\bf 1}^9\oplus{\bf
  3}^{27}}_{\textrm{untwisted}}\oplus\underbrace{{\bf 1}^{18}\oplus{\bf 3}^{54}}
_{\textrm{twisted}}.
\end{equation}
The (2,2) moduli are ${\bf 1}^3\oplus{\bf 3}^9\oplus{\bf 1}^6$ as
expected.  

Now we move away from the Gepner point but remain mirror to the {\em
  orbifold\/} by 
adding $\Qb_O$.  In the untwisted sector, as
noted above, the rank of the map between $\SU(3)$ singlet states
increases from 27 to 33.  This is the Higgs mechanism breaking the
enhanced gauge symmetry $\SU(3)\times\GU(1)^6$ to $\SU(3)$ by removing a
vector multiplet and a chiral multiplet for each of the six broken generators of
the gauge group.  In the twisted sector we find that
the rank of the map on singlets increases from 36 to 48.  In the
$(1;1)$ sector the one-dimensional kernel at $\qb=\ff12$ is now spanned
by $\sum_{ijk}\epsilon^{ijk}\rho_{xi}\gammab_{xj}\gammab_{xk}|1;1\ra$.  The
cohomology thus contains 9 vector multiplets in the representation 
${\bf 1}\oplus{\bf 8}$
and 252 chiral multiplets in the representation 
\begin{equation}
\underbrace{{\bf 1}^3\oplus{\bf
  3}^{27}}_{\textrm{untwisted}}\oplus\underbrace{{\bf 1}^{6}\oplus{\bf
  3}^{54}}_{\textrm{twisted}}
\ ,
\end{equation}
as was found in \cite{DHVW:}.

Finally we blow-up the orbifold in the mirror by adding $\Qb_D$. Now
the enhanced gauge symmetry is completely broken.  In the untwisted sector we saw
that the rank of $\Qb$ grows by two.  This is again the Higgs
mechanism for the Cartan elements of the $\SU(3)$ gauge symmetry, and
it is accompanied by a corresponding change in rank by one in the map
$\qb=-\ff32\to\qb=-\ff12$ in each twisted sector.  In addition, we find
that the rank of the map $\qb=-\ff12\to\qb=\ff12$ increases from 62 to 80,
indicating that 108 chiral multiplets are lifted by spacetime superpotential
interactions ($F$-terms).  This leaves a total of 136 $E_6$ neutral
chiral multiplets at generic points in the moduli space.

Comparing these numbers to the results in the previous section we find
that a worldsheet instanton generated spacetime superpotential leads
to mass terms for 108 of the perturbatively massless scalars found by
the large-radius analysis.  


\section{Summary and Discussion}

It is surprising that $h^1(\End(T))$ is so resilient to instanton
corrections. It would seem that there are mechanisms beyond those of
\cite{Silverstein:1995re,Beasley:2003fx} that protect these states
from acquiring masses away from the large radius limit. From the
examples we have considered, it is tempting to conjecture that a good candidate 
for a class of models in which there are
no corrections is that of hypersurfaces in toric
varieties. Typically these have contributions to $h^1(\End(T))$ above
the zeroth row of the spectral sequence and, in all cases we have
considered, their masses are not affected by instantons.  

The fact that there seems to be a large class of models without
corrections should lead to some interesting mathematics. The agreement
between the cohomology of the complexes of the \LG\ picture and the
spectral sequence of the $\sigma$-model is not yet understood. One may
also be able to make curious statements such as $h^1(\End(T))$ being
equal for mirror pairs of \CY\ threefolds.

On the other hand, we now have an example where instantons {\em do\/}
give mass to $H^1(\End(T))$ modes. An interesting next step will be to
enumerate these instantons more precisely from the conformal field
theory computation.

In the case of the $Z$-manifold, we find that of the 252 massless $E_6$ singlet 
fields in the $Z$-orbifold spectrum, 8 acquire $D$-term masses via the Higgs
mechanism and 108 acquire $F$-term masses via a spacetime
superpotential as we blow up the quotient singularities.  Removing the
36 $(2,2)$ moduli, this leaves precisely 100 massless singlets
corresponding to bundle deformations at a generic point in the
(30-dimensional) moduli space.  Comparing this to the geometric
computation in the large-radius limit we find again that of the 208
singlet fields counted by $h^1(\End(T))$ 108 acquire masses via 
instanton-induced superpotential terms.   It is natural to conjecture
that one can identify the full 208-dimensional space of massless
states at the orbifold locus with $H^1(\End(T))$ and that the two sets
of $F$-terms are associated to the same superpotential.

This counting argument suggests that even in this case where
instanton contributions are nonzero there is a 
McKay-like correspondence for local contributions to $h^1(\End(T))$
form an orbifold resolution. This would imply that there is some local
picture for noncompact \CY\ threefolds where instantons can be shown to cancel.
Clearly further investigation of this would be interesting.

\section*{Acknowledgments}

We thank I.~Melnikov and E.~Miller for useful discussions.
This work was partially supported by NSF grants DMS--0606578 and
DMS--0905923.  Any opinions, findings, and conclusions or
recommendations expressed in this material are those of the authors
and do not necessarily reflect the views of the National Science
Foundation.

\appendix

\section{$h^1(\End(T))$ for the Z-manifold}
$Z$ is the resolution of $T^6/\Z_3$ at the 27 fixed points of the
orbifold action. We may write the 2-torus as a cubic in $\P^2$. Hence
$Z$ can be written as a complete intersection of three ``cubics'' in
$V$ where $V$ is the crepant desingularization of
\begin{equation}
  V' = \frac{\P^2\times \P^2\times \P^2}{\Z_3}.
\end{equation}
$V$ is a toric variety as follows. The 1-dimensional rays of the
associated fan are given by the rows of the matrix
\begin{equation}\left(\begin{smallmatrix}1&
    1& {-2}& 1& 1&
    0\\
    {-1}& {-1}& {2}& {-1}& {-1}&
    {-1}\\
    0& 0& 0& 0& 0&
    1\\
    0& 0& 0& {-1}& {-1}&
    0\\
    0& 0& 0& 1& 0&
    0\\
    0& 0& 0& 0& 1&
    0\\
    {-1}& {-1}& {-1}& 0& 0&
    0\\
    1& 0& 1& 0& 0&
    0\\
    0& 1& 0& 0& 0&
    0\\
    0& 0& {-1}& 0& 0&
    0\\
\end{smallmatrix}\right)
\end{equation}
The homogeneous coordinates $(x_0,\ldots,x_9)$ are associated to these
rays. The Stanley-Reisner ideal is given by
\begin{equation}
I = {\langle{x}_{0} {x}_{1} {x}_{2},{x}_{3} {x}_{4} {x}_{5},{x}_{0} {x}_{3}
  {x}_{6},{x}_{6} {x}_{7} {x}_{8},{x}_{1} {x}_{2} {x}_{9},{x}_{4}
  {x}_{5} {x}_{9},{x}_{7} {x}_{8} {x}_{9}\rangle}.
\label{eq:SR-Z}
\end{equation}
$Z$ is now the complete intersection in $V$ associated to the ideal
$\langle f^0,f^1,f^2\rangle$ where
\begin{equation}
\begin{split}
f^0 &= {x}_{1}^{3}+{x}_{2}^{3}+{x}_{0}^{3} {x}_{9}\\
f^1 &= {x}_{4}^{3}+{x}_{5}^{3}+{x}_{3}^{3} {x}_{9}\\
f^2 &= {x}_{7}^{3}+{x}_{8}^{3}+{x}_{6}^{3} {x}_{9}.
\end{split}
\end{equation}

We have a representation of the sheaf $\End(T_Z)$ in terms of a
complex of sums of line bundles on $Z$ given by (\ref{eq:HomTT}). It
is more convenient for computational purposes if everything is written
in terms of line bundles over the ambient toric variety $V$.

To see how to do this let us consider the much simpler case of the
tangent bundle of the quintic 3-fold. This is given as the complex
\begin{equation}
\xymatrix@1{
\O_X\ar[r]^-{x_i}&\O_X(1)^{\oplus5}\ar[r]^-{\partial_if}&\O_X(5).}
\label{eq:Tquin}
\end{equation}
Let $\O$ denote the structure sheaf of the ambient $\P^4$. $\O_X$ is
then equivalent, in the derived category, to
\begin{equation}
\xymatrix@1{
\O(-5)\ar[r]^-f&\O.}
\end{equation}
Thus, by mapping cones, the complex given by the last two terms of
(\ref{eq:Tquin}) is given by
\begin{equation}
\xymatrix@1@C=20mm{
\O(-4)^{\oplus5}\ar[r]^{\left(\begin{smallmatrix}
-fI\\\partial_if\end{smallmatrix}\right)}
&{\begin{matrix}\O(1)^{\oplus5}\\\oplus\\
\O\end{matrix}}\ar[r]^{\left(\begin{smallmatrix}
\partial_if&f\end{smallmatrix}\right)}&\O(5).}
\label{eq:Tquin2}
\end{equation}
The complete complex (\ref{eq:Tquin}) is then given by the mapping
cone of a map from complex representing $\O_X$ to the complex 
(\ref{eq:Tquin2}). This morphism is given by the chain map
\begin{equation}
\xymatrix@C=20mm{
\O(-5)\ar[r]^-f\ar[d]^{x_i}&\O\ar[d]^-{\left(\begin{smallmatrix}
-x_i\\5\end{smallmatrix}\right)}\\
\O(-4)^{\oplus5}\ar[r]^{\left(\begin{smallmatrix}
-fI\\\partial_if\end{smallmatrix}\right)}
&{\begin{matrix}\O(1)^{\oplus5}\\\oplus\\
\O\end{matrix}}\ar[r]^{\left(\begin{smallmatrix}
\partial_if&f\end{smallmatrix}\right)}&\O(5).
}
\end{equation}
Note, in particular, the need for the ``5'' to produce a chain map. 
The tangent sheaf of the quintic can thus be represented by the chain
complex
\begin{equation}
\xymatrix@1@C=20mm{
\O(-5)\ar[r]^-{\left(\begin{smallmatrix}
-f\\x_i\end{smallmatrix}\right)}&
{\begin{matrix}\O\\\oplus\\\O(-4)^{\oplus5}\end{matrix}}
\ar[r]^-{\left(\begin{smallmatrix}-x_i&-fI\\5&\partial_if\end{smallmatrix}
\right)}&
{\begin{matrix}\O(1)^{\oplus5}\\\oplus\\
\O\end{matrix}}\ar[r]^{\left(\begin{smallmatrix}
\partial_if&f\end{smallmatrix}\right)}&
\O(5).}
\end{equation}

We
need to extend this construction to complete intersections. To this
end we may prove the following theorem analogously to the above.
\begin{theorem}
Consider a commutative diagram of sheaves on $V$ of the following form:
\begin{equation}
\xymatrix{
\ldots\ar[r]&A^{n-1}\ar[r]^{a_{n-1}}\ar[d]^{h_{n-1}}&A^{n}\ar[r]^{a_n}\ar[d]^{h_{n}}&A^{n+1}\ar[d]^{h_{n+1}}\ar[r]&\ldots\\
\ldots\ar[r]&B^{n-1}\ar[r]^{b_{n-1}}\ar[d]^{g_{n-1}}&B^{n}\ar[r]^{b_n}\ar[d]^{g_{n}}&B^{n+1}\ar[d]^{g_{n+1}}\ar[r]&\ldots\\
\ldots\ar[r]&C^{n-1}\ar[r]^{c_{n-1}}\ar[d]^{f_{n-1}}&C^{n}\ar[r]^{c_n}\ar[d]^{f_{n}}&C^{n+1}\ar[d]^{f_{n+1}}\ar[r]&\ldots\\
\ldots\ar[r]&D^{n-1}\ar[r]^{d_{n-1}}&D_{n}\ar[r]^{d_n}&D^{n+1}\ar[r]&\ldots
}
\end{equation}
satisfying
\begin{enumerate}
\item Each column is exact except it has cohomology $L^n$ supported on $Z$
  in the last position.
\item the composition of two horizontal maps is zero when restricted
  to $Z$.
\end{enumerate}
The complex
$L^\bullet$ is quasi-isomorphic to the complex with terms
$A^{n+3}\oplus B^{n+2}\oplus C^{n+1}\oplus D^n$ and differentials

\begin{equation}
\begin{pmatrix}
a_{n+3}&(-1)^n\alpha_{n+2}&\beta_{n+1}&(-1)^n\gamma_n\\
(-1)^n h_{n+3}&b_{n+2}&(-1)^n\delta_{n+1}&\epsilon_n\\
0&(-1)^n g_{n+2}&c_{n+1}&(-1)^n\zeta_n\\
0&0&(-1)^n f_{n+1}&d_n
\end{pmatrix}:
\begin{pmatrix} A^{n+3}\\B^{n+2}\\C^{n+1}\\D^n\end{pmatrix}
\longrightarrow
\begin{pmatrix} A^{n+4}\\B^{n+3}\\C^{n+2}\\D^{n+1}\end{pmatrix}
\end{equation}
where
\begin{equation}
\begin{split}
\zeta_n&=f^{-1}_{n+2}d_{n+1}d_n\\
\delta_n &= g^{-1}_{n+2}(c_{n+1}c_n - \zeta_nf_n)\\
\epsilon_n &= g^{-1}_{n+3}(c_{n+2}\zeta_n - \zeta_{n+1}d_n)\\
\alpha_n &= h^{-1}_{n+2}(b_{n+1}b_n - \delta_ng_n)\\
\beta_n &= h^{-1}_{n+3}(b_{n+2}\delta_n - \delta_{n+1}c_n + \epsilon_n
           f_n)\\
\gamma_n &= h^{-1}_{n+4}(b_{n+3}\epsilon_n - \delta_{n+2}\zeta_n + \epsilon_{n+1}d_n)
\end{split}
\end{equation}
\end{theorem}

Let the multi-degree of the equation $f_i$ be denoted $\mathbf{Q}_i$.
 The sheaf
$\O_Z$ can be resolved in terms of the Koszul complex:
\begin{equation}
\xymatrix@1{
  \O(-\sum_i\mathbf{Q}_i)\ar[r]^-h&
  \bigoplus_i\O(-\sum_{j\neq i}\mathbf{Q}_j)\ar[r]^-g&
  \bigoplus_i\O(-\mathbf{Q}_i)\ar[r]^-f&
  \O\ar[r]&\O_Z,}
\end{equation}
where $\O$ is the structure sheaf of $V$. 

Applying the theorem to this and the complex (\ref{eq:HomTT}) yields a
rather messy complex of line bundles on $V$ representing $\End(T)$ for
the $Z$-manifold. The explicit form is too large to give here and we
will write it more concisely as
\begin{equation}
\xymatrix@1{
  \cF^{-5}_{(12)}\ar[r]&\cF^{-4}_{(106)}\ar[r]&\cF^{-3}_{(371)}\ar[r]&
  \cF^{-2}_{(667)}\ar[r]&\cF^{-1}_{(667)}\ar[r]&\cF^{0}_{(371)}\ar[r]&
  \cF^{1}_{(106)}\ar[r]&\cF^{2}_{(12)}}
\label{eq:big1}
\end{equation}
where each term is a sum of line bundles on $V$ where the rank is
denoted by the subscript.

Now, to compute the cohomology we have a spectral sequence $E_1^{p,q}=
H^q(\cF^p)$. This yields
\begin{equation}
\def\objectstyle{\scriptstyle}
\begin{matrix}\vspace{20mm}\\E_1^{p,q}:\end{matrix}
\begin{xy}
\xymatrix@C=8mm@R=6mm{\scriptstyle
  60\ar[r]&100\ar[r]&35\\
  &&&&&0\\
  &104\ar[r]&432\ar[r]&510\ar[r]&234\ar[r]&24\\
  &&48\ar[r]&258\ar[r]&258\ar[r]&48\\
  &&24\ar[r]&234\ar[r]&510\ar[r]&432\ar[r]&104\\
  &&&&&0\\
  &&&&&35\ar[r]&100\ar[r]&60
}
\save="x"!LD+<-3mm,0pt>;"x"!RD+<0pt,0pt>**\dir{-}?>*\dir{>}\restore
\save="x"!LD+<68mm,-3mm>;"x"!LU+<68mm,-2mm>**\dir{-}?>*\dir{>}\restore
\save!CD+<0mm,-4mm>*{p}\restore
\save!UL+<66mm,0mm>*{q}\restore
\save="x"!LD+<-2mm,65mm>;"x"!LD+<92mm,0mm>**\dir{.}\restore
\end{xy} \label{eq:Tss1}
\end{equation}
where each number represents the dimension of the space at each
position. The dotted line shows the terms that will ultimately
contribute to $h^1(\End(T))$. Now we need to compute the maps above
and take cohomology to proceed to $E_2^{p,q}$.

To do this we follow the method given in \cite{meMP:singlets}. We may
regard the spectral sequence as having arisen from a double complex
with the given complex maps in the horizontal direction and \v Cech
cohomology in the vertical direction. 

Let $B=(m_1,m_2,\ldots,m_l)$, where $m_i$ are monomials, be the
irrelevant ideal which is the Alexander dual \cite{MS:combcomm} of the
Stanley--Reisner ideal (\ref{eq:SR-Z}). An open cover of $V$ that can
be used to compute \v Cech cohomology is then given by
the set of $U_i=V-Z(m_i)$. This is equivalent to a local cohomology
computation as explained in \cite{EMS:ToricCoh}. This local cohomology
approach to computing the cohomology of line bundles has also been
explored in the context of string theory in
\cite{Herzog:2005sy,me:toricD,Blumenhagen:2010pv}. We refer to the
appendix of \cite{meMP:singlets} for a review of the ideas required here.

An element of the underlying double complex is given by a \v Cech
cochain which is a collection of Laurent monomials
$\{c_{j_1,j_2,\ldots}\}$ each of which takes values in the
localization $R_{m_{j_1},m_{j_2},\ldots}$. This structure simplifies a
little once we go to the $E_1$ stage by taking vertical
cohomology. Each cochain must be coclosed and so we require exact
cancellations on certain multiple overlaps of open patches
$U_i$. Suppose we fix one of the $c_{j_1,j_2,\ldots}$'s to be a fixed
monomial of homogeneous coordinates. In straight-forward cases, such a
cancellation tends to require that the other $c_{j_1,j_2,\ldots}$'s
are given (perhaps up to some fixed constant) by the exact same
monomial. Thus we specify a given dimension of the space of cohomology
by specifying a particular monomial. In other words, computing the
dimension of the cohomology amounts to counting Laurent monomials of a
certain form. 

Whether this simple counting method works depends on the finely-graded
Betti numbers $\Tor_i^R(I,\C)_\alpha$ as in Corollary 3.1 of
\cite{Must:LocalCoh}. They are required all be 0 or 1. While there are
certainly counterexamples where these Betti numbers are bigger than
one, such examples are combinatorially quite complicated and,
fortunately, the $Z$-manifold does not fall into this class.

The upshot of all this is that we can represent entries in $E_1^{p,q}$
simply by Laurent monomials (rather than collections of monomials
associated to intersections of patches). The combinatorics of counting
such monomials has been discussed in
\cite{EMS:ToricCoh,Blumenhagen:2010pv}. What is even nicer is that it
easily follows that the horizontal maps in the $E_1^{p,q}$ stage of
the spectral sequence are given by the obvious maps between monomials
inherited from the original complex (\ref{eq:big1}). An exercise in
Macaulay 2 programming then yields:

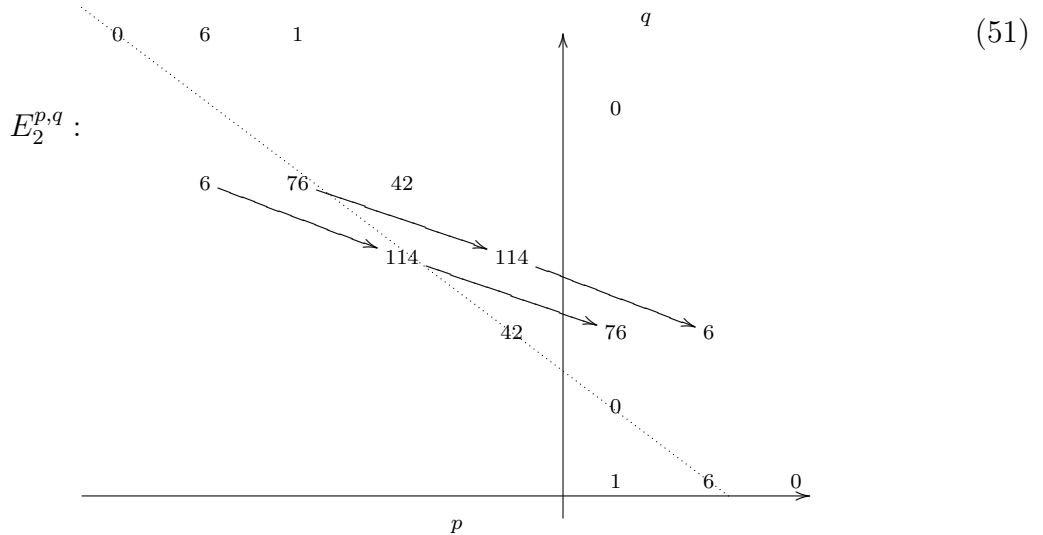
\begin{equation}
\def\objectstyle{\scriptstyle}
\begin{matrix}\vspace{20mm}\\E_2^{p,q}:\end{matrix}
\begin{xy}
\xymatrix@C=8mm@R=6mm{
  0&6&1\\
  &&&&&0\\
  &6\ar[rrd]&76\ar[rrd]&42\\
  &&&114\ar[rrd]&114\ar[rrd]\\
  &&&&42&76&6\\
  &&&&&0\\
  &&&&&1&6&0
}
\save="x"!LD+<-3mm,0pt>;"x"!RD+<0pt,0pt>**\dir{-}?>*\dir{>}\restore
\save="x"!LD+<61mm,-3mm>;"x"!LU+<61mm,-2mm>**\dir{-}?>*\dir{>}\restore
\save!CD+<0mm,-4mm>*{p}\restore
\save!UL+<72mm,0mm>*{q}\restore
\save="x"!LD+<-3mm,65mm>;"x"!LD+<83mm,0mm>**\dir{.}\restore
\end{xy} \label{eq:Tss2}
\end{equation}

We now need to compute the $d_2$ maps shown in the above
diagram. These are computed following the staircases of maps as
described in \cite{BT:}. We need to go right-down-right. An example of
this was described completely explicitly in \cite{meMP:singlets} and so we
will be brief here. The composition of two right maps obviously gives
zero in the underlying complex (\ref{eq:big1}). However, the downwards
map serves to scramble some of the signs in the map because of the
sign choices in \v Cech cohomology. Thus, the $d_2$ maps need not be
zero. Fortunately most of the entries in the matrices representing
$d_2$ can be shown to be zero because of the large number of zeroes in
the horizontal maps. The result is:

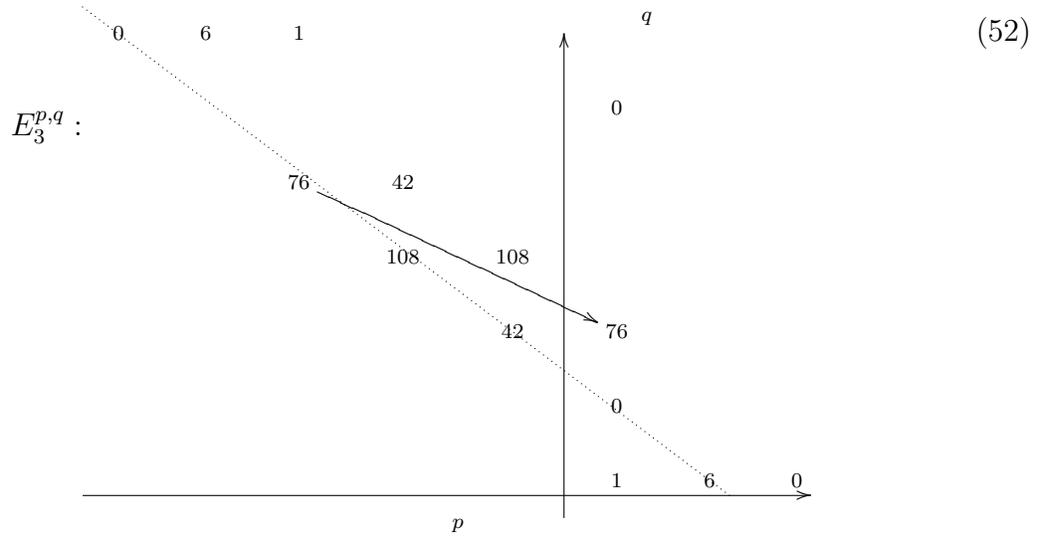
\begin{equation}
\def\objectstyle{\scriptstyle}
\begin{matrix}\vspace{20mm}\\E_3^{p,q}:\end{matrix}
\begin{xy}
\xymatrix@C=8mm@R=6mm{
  0&6&1\\
  &&&&&0\\
  &&76\ar[rrrdd]&42\\
  &&&108&108\\
  &&&&42&76\\
  &&&&&0\\
  &&&&&1&6&0
}
\save="x"!LD+<-3mm,0pt>;"x"!RD+<0pt,0pt>**\dir{-}?>*\dir{>}\restore
\save="x"!LD+<61mm,-3mm>;"x"!LU+<61mm,-2mm>**\dir{-}?>*\dir{>}\restore
\save!CD+<0mm,-4mm>*{p}\restore
\save!UL+<72mm,0mm>*{q}\restore
\save="x"!LD+<-3mm,65mm>;"x"!LD+<83mm,0mm>**\dir{.}\restore
\end{xy} \label{eq:Tss3}
\end{equation}

Mercilessly, the spectral sequence is not done with us yet and we need to
compute one $d_3$ map. This can be tackled using the same method as we
used for $d_2$. The result is:

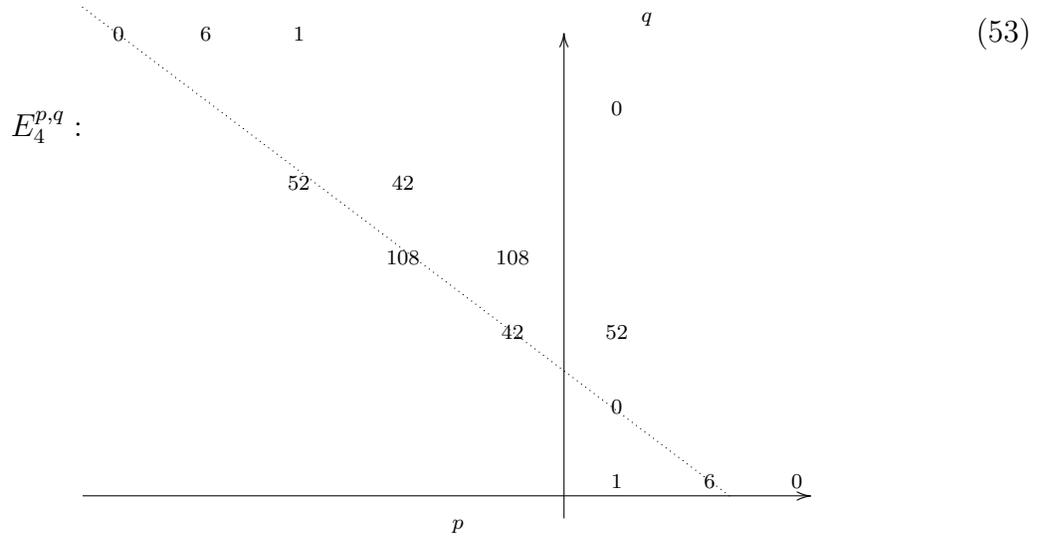
\begin{equation}
\def\objectstyle{\scriptstyle}
\begin{matrix}\vspace{20mm}\\E_4^{p,q}:\end{matrix}
\begin{xy}
\xymatrix@C=8mm@R=6mm{
  0&6&1\\
  &&&&&0\\
  &&52&42\\
  &&&108&108\\
  &&&&42&52\\
  &&&&&0\\
  &&&&&1&6&0
}
\save="x"!LD+<-3mm,0pt>;"x"!RD+<0pt,0pt>**\dir{-}?>*\dir{>}\restore
\save="x"!LD+<61mm,-3mm>;"x"!LU+<61mm,-2mm>**\dir{-}?>*\dir{>}\restore
\save!CD+<0mm,-4mm>*{p}\restore
\save!UL+<72mm,0mm>*{q}\restore
\save="x"!LD+<-3mm,65mm>;"x"!LD+<83mm,0mm>**\dir{.}\restore
\end{xy} \label{eq:Tss4}
\end{equation}

Finally the spectral sequence degenerates and so
\begin{equation}
  h^1(\End(T)) = 208.
\end{equation}


\end{document}